\documentclass[a4paper]{article}
\usepackage{a4wide}
\usepackage[english]{babel}
\usepackage{graphicx}
\usepackage{amsmath,amsthm}
\usepackage{amsfonts}

\newcommand{\R}{\mathbb{R}}
\newcommand{\Z}{\mathbb{Z}}

\newcommand{\cylindre}{{\cal C}_r}

\renewcommand{\L}{L}

\numberwithin{equation}{section}
\theoremstyle{plain}

\def\eff{{\rm eff}}

\def\demi{\frac{1}{2}}
\def\demi{\frac{1}{2}}

\usepackage{color} 

\begin{document}

\noindent 
\begin{center}
\textbf{\large Rigorous perturbation theory versus variational methods
  in the spectral study of carbon nanotubes}
\end{center}

\begin{center}
May 1st, 2007
\end{center}

\vspace{0.5cm}

\noindent 

\begin{center}
\textbf{ Horia D. Cornean\footnote{Dept. of Mathematical Sciences, 
    Aalborg
    University, Fredrik Bajers Vej 7G, 9220 Aalborg, Denmark; e-mail:
    cornean@math.aau.dk}, Thomas G. Pedersen\footnote{Dept. Phys. 
and Nanotech., 
    Aalborg
    University, 9220 Aalborg, Denmark; e-mail:
    tgp@physics.aau.dk}, 
Benjamin Ricaud \footnote{Centre de Physique Th\'eorique UMR 6207 -
  Unit\'e Mixte de Recherche du CNRS et des Universit\'es
  Aix-Marseille I, 
Aix-Marseille II et de l'universit\'e du Sud Toulon-Var - Laboratoire
affili\'e \`a la FRUMAM, Luminy Case 907, F-13288 
Marseille Cedex 9 France; 
e-mail: ricaud@cpt.univ-mrs.fr}}
     
\end{center}

\vspace{0.5cm}

\begin{abstract} 
Recent two-photon photo-luminescence experiments 
give accurate data for the ground and first excited excitonic energies at different nanotube radii. In this paper we compare
the analytic approximations proved in \cite{CDR}, with a standard variational approach. We show an excellent
agreement at sufficiently small radii. 

\end{abstract} 
\section{Introduction}

Recent experimental results on carbon nanotubes using two photon
photo luminescence~\cite{Sc02}, \cite{Sc05}, \cite{PRB05} reveal the
energy levels of the excitons, especially the ground and first excited
states, and point out the dependence of these energies on the radius
of the nanotube. As it is, this technique appears to be a promising
way to sort out nanotubes. But on the other hand, theoretical results
seem to require heavy ab initio calculation like
in~\cite{PRB05}~\cite{PRB05-2}, to cite the most recent, in order to
find the absorption peaks due to excitons. Nevertheless, a simple
formula for the optical response based on excitons levels~\cite{HK}
can give a good approximation. It has been pointed
out~\cite{PRB87}~\cite{PRB91} that the exciton binding energy in
quantum wires depends on the width of the wires by a relatively simple
relation. This property is also valid for nanotubes~\cite{P1}. In the
first part of the paper, we outline a rigorous justification for this
latter fact and write a simple analytic formula for the energy
levels of the exciton depending on the radius of the tube, based on
the paper~\cite{CDR}. In the second part we compare our results with a
variational numerical method and show a very good agreement between them.

\section{The exciton model}
As first suggested in~\cite{P1}, we deal with Wannier excitons (a
rigorous justification of this procedure is in preparation
\cite{B}). We take as configuration space 
a cylinder of radius $r$ and infinite length, space denoted by $\cylindre=\R\times r S^1$, $S^1$ being the
 unit circle. The coordinates on the cylinder are $(x,y)\in(\R\times r
S^1)$ where $x$ is the variable along the tube axis and $y$ is the
transverse coordinate.

The two quasi-particles live in the Hilbert space
$\L^2(\cylindre\times \cylindre)$. We formally consider the Hamiltonian
\begin{equation}\label{pricipessa}
\bar{H}^r=-\hbar^2\left(\frac{\Delta_{x_1}}{2m_1}+\frac{\Delta_{x_2}}{2m_2}+\frac{\Delta_{y_1}}{2m_1}+
\frac{\Delta_{y_2}}{2m_2}\right)-V^r(x_1-x_2,y_1-y_2),
\end{equation}
where
\begin{equation}\label{zaculomb}
V^r(x,y):=\frac{-e_1 e_2}{\varepsilon\sqrt{x^2+4r^2\sin^2\left(\frac{y}{2r}\right)}}
\end{equation}
$(x_i,y_i)$ are the coordinates on the cylinder of the two charged
particles, $m_i$ their masses, and $e_i$ their charges. Here
$\varepsilon$ is the electric permittivity of the material. The
potential $V^r$ is the three dimensional Coulomb potential simply
restricted to the cylinder. We justify the expression of $V^r$ by
Pythagoras's theorem. The cylinder is embedded in $\R^3$. The distance $\rho$ from one particle to the other in $\R^3$ is:
$$\rho^2=(x_1-x_2)^2+4r^2\sin^2\left(\frac{y_1-y_2}{2r}\right)$$
where $|2r\sin\left(\frac{y_1-y_2}{2r}\right)|$ is the length of the
chord joining two points of coordinate $y_1$ and $y_2$ on the circle.

\subsection{"Separation of the center of mass"}
Due to the restrictions imposed by the cylindrical geometry, the usual
separation of the center of mass does not work here. 

The standard separation with Jacobi coordinates only works for the
longitudinal variable, and introducing $M:=m_1+m_2$ and 
$\mu:=m_1 m_2/(m_1+m_2)$, we denote $X:=(m_1x_1+m_2x_2)/M$ and
$x:=x_1-x_2$. For the transverse variable, 
we change to atomic coordinates $Y=y_2$ and $y=y_1-y_2$. Let us also
define the effective Rydberg $Ry^*=\mu e^4/2\hbar^2\varepsilon^2$ 
and Bohr radius $a_B^*=\hbar^2\varepsilon/\mu e^2$, where we set
$e=e_1=e_2$. By a scaling, the new energy and radius will be 
expressed in multiple of these units. 
This gives us the Hamiltonian:
\begin{eqnarray}
H^r&=&-\frac{1}{M}\partial_X^2
-\frac{1}{m_2}\partial_Y^2-\frac{1}{\mu}\partial_x^2
-\frac{1}{\mu}\partial_y^2 +
\frac{2}{m_2}\partial_y\partial_Y- 2V^r(x,y).\nonumber
\end{eqnarray}
First, we can separate the partial center of mass with coordinate $X$. Second, 
since on the $Y$ variable there are periodic boundary conditions, let us consider the orthonormal basis of eigenvectors of $-\partial_Y^2$, 
$$-\partial_Y^2=\sum_{n=-\infty}^{\infty} E_n^r |\chi_n^r\rangle\langle\chi_n^r| 
$$
where
$$\chi_n^r(Y)=\frac{1}{\sqrt{2\pi r}}e^{in\frac{Y}{r}}\mbox{ and
}E_n^r=\frac{n^2}{r^2},\> n\in \Z .
$$
One can see that for small radii, the separation between different
transverse levels of energy is high. A recent theoretical study based
on ab initio calculations~\cite{PRB05} shows that the probability
density of the exciton is constant along the circumference. So it is
reasonable to assume that the radius is so small that the system stays
in the ground state of $-\partial_Y^2$, where $n=0$ and the density of
probability is constant along the circumference. (Note that this has
been rigorously proved in \cite{CDR}). As a consequence, we can approximate the eigenfunctions $\psi$ of $H$ as:
$$\psi(x,y,Y)=\phi(x,y)\cdot\chi_0^r(Y)=\phi(x,y)\cdot \frac{1}{\sqrt{2\pi r}}.
$$
After this restriction to the lowest transverse mode, we only have to study the following operator:
\begin{eqnarray}
\widetilde{H}^r&=&-\frac{1}{\mu}\partial_x^2 -\frac{1}{\mu}\partial_y^2 - 2V^r(x,y)\nonumber,
\end{eqnarray}
which is two dimensional and acts on $L^2(\cylindre)$.

\subsection{An effective one dimensional operator for the low lying spectrum}

It will turn out that the limit $r\to 0$ is too singular and
$\widetilde{H}^r$ does not have a ``nice'' limit. It is suitable at
this point to introduce the quadratic form associated to
$\widetilde{H}^r$, defined on the Sobolev space $\mathcal{H}^1(\mathcal{C}_r)$:
\begin{align}
t_{\widetilde{H}}(\psi,\phi)=\frac{1}{\mu}\langle\partial_x
\psi,\partial_x\phi\rangle+
\frac{1}{\mu}\langle \partial_y \psi,\partial_y\phi\rangle-2
\langle\sqrt{V^r(x,y)}\psi,\sqrt{V^r(x,y)} \phi\rangle.\nonumber
\end{align}
 Reasoning as in the previous subsection, a good approximation of the behavior
 of eigenfunctions along the transverse variable when the radius
 is small is given by the ground state of the free Laplacian with
 periodic boundary conditions. If we restrict the above quadratic form
 to functions of the type:
\begin{equation}
\phi(x,y)=\varphi(x)\cdot\frac{1}{\sqrt{2\pi r}},
\end{equation}
we have:
\begin{align*}
& t_{\widetilde{H}}(\phi,\phi)= \frac{1}{2\pi r\mu}\langle \partial_x \varphi,\partial_x
\varphi\rangle - \frac{1}{\pi r}\int_{-\infty}^{\infty}\int_{-\pi r}^{\pi r}\frac{1}{\sqrt{x^2+4r^2\sin^2\frac{y}{2r}}}\varphi(x)\overline{\varphi}(x)dydx\\
&=\frac{1}{\mu}\int_{-\infty}^{\infty}\overline{(\partial_x\varphi)}(x)(\partial_x\varphi)(x)dx-2 \int_{-\infty}^{\infty}V_{\rm eff}^r(x)\varphi(x)\overline{\varphi}(x)dx
\end{align*}
where
$$V_{\rm eff}^r(x)=\frac{1}{2\pi r}\int_{-\pi r}^{\pi r}\frac{1}{\sqrt{x^2+4r^2\sin^2\frac{y}{2r}}}dy.
$$
For the sake of simplicity, we will put $\mu=1$. Let us introduce what will be our
effective one-dimensional comparison operator: 
\begin{equation}\label{hasefectiv}
H_{\eff}^r:=-\frac{d^2}{dx^2}-2V_{\rm eff}^r(x).
\end{equation}
We now have reduced our problem of two particles on a cylinder to a
one dimensional problem describing a particle interacting with an
external potential. A complete and detailed mathematical justification of these steps can be found in
\cite{CDR}. 

\subsection{The one dimensional Coulomb Hamiltonian}
\subsubsection{Approximation of $H_{\rm eff}^r$}
We now try to approximate as accurate as possible the eigenfunctions
and eigenvalues of $H_{\rm eff}^r$ when the radius is small. 
Using the notation $Y_r(x)=\frac{1}{\sqrt{x^2+4r^2}}$, 
it was shown in \cite{CDR} that if $r\le1$ and $f,g$ are two smooth functions
\begin{equation}
\langle f,V_{\rm eff}^r g\rangle=
\langle f,Y_r g\rangle+\ln(4)f(0)\overline{g(0)} +O(r^{4/9})||f||_{\mathcal{H}^1(\mathbb{R})}||g||_{\mathcal{H}^1(\mathbb{R})}.
\end{equation}
One can recognize in $Y_r$ a form of the regularized one dimensional Coulomb
potential. In the following, our approach will be quite similar to the
one of Loudon~\cite{L} in the sense that we let the regularizing
parameter tend to zero. Note though that our parameter has a clear
physical interpretation being given by the radius of the nanotube, and it is not
just an artifact as in Loudon's case. Hence this will give us the
exciton energies as functions of $r$, and allow us to estimate the
errors we make using this approximation. 

Define the quadratic form $C_0(.,.)$ by\footnote[1]{Notice that this
  definition of $C_0$ differs from the one of~\cite{CDR} 
in which there is an additional term: 
$2\ln 2 |f(0)|^2=\int_{-\infty}^0 \ln 2 \cdot
(|f|^2)'(x)dx+\int_{0}^{\infty} \ln 2 \cdot (|f|^2)'(x)dx$. 
This term is here put together with the other term depending on the function at zero.} 
\begin{align}\label{formuleC_0}
 C_0(f,f)&: =-\int_0^{\infty} \ln(x)\cdot (|f|^2)'(x)dx+\int_{-\infty}^0 \ln(-x)\cdot(|f|^2)'(x)dx\nonumber\\
&=\int_{-\varepsilon}^0 \ln(-x)\cdot (|f|^2)'(x)dx-\int_0^{\varepsilon} \ln(x)\cdot (|f|^2)'(x)dx\nonumber\\
&+\ln(\varepsilon)\left(|f(\varepsilon)|^2+|f(-\varepsilon)|^2\right)+\int_{\R\backslash\left[-\varepsilon,\varepsilon\right]} \frac{1}{|x|}\cdot |f(x)|^2dx.
\end{align}
Note the last equality is obtained by an integration by parts, and
holds for all $\varepsilon>0$; also note the
appearance of a Coulomb potential in one dimension. Again from~\cite{CDR} we have
\begin{equation}
\langle f,Y_rg\rangle = -2\ln(r)f(0)\overline{g(0)}+C_0(f,g)+ O(r^{4/9})||f||_{\mathcal{H}^1(\mathbb{R})}||g||_{\mathcal{H}^1(\mathbb{R})}.
\end{equation}
Now let us define a new ``potential'' via a quadratic form
$$\langle f,V_Cg\rangle:=-2\ln(\frac{r}{2})f(0)\overline{g(0)}+C_0(f,g),
$$
which is close to $V_{\eff}^r$ when $r$ is small and is exactly the Coulomb potential when we look away from the origin.
We will see in the following that this particular potential gives a
solvable eigenvalue problem for the associated Hamiltonian.
\subsubsection{Boundary conditions of the Coulomb Hamiltonian}
The operator $H_C$ we will now consider is given by its associated
quadratic form defined on $\mathcal{H}^1(\mathbb{R})$:
$$t_C(\psi,\phi):=\int_{\R}\overline{\phi'(x)}
\psi'(x)dx+2\left(2\ln(\frac{r}{2})\overline{\phi(0)}\psi(0)-C_0(\phi,\psi)\right).
$$
 One can recognize the kinetic energy in the first term and the
 potential $V_C$ in the second term. The general theory of closed symmetric
 quadratic forms gives us the existence of an 
associated operator $H_C$ defined by
\begin{equation}\label{centralstation}
t_C(\phi,\psi)=\langle\phi,H_C\psi\rangle,
\end{equation}
whenever $\psi$ is in the domain of $H_C$. The difficulty here is that
$V_C$ is not a usual Schr\"odinger-type multiplication potential, but
due to various Sobolev embeddings it turns out that if $\psi$ is in
the domain of $H_C$ then $\psi''$ is square
integrable outside the origin, and we still have:
\begin{equation}\label{eqHc}
(H_C\psi)(x)=-\psi''(x)-\frac{2\psi(x)}{|x|},\quad x\neq 0.
\end{equation}
Now if $\psi$ is an eigenfunction of $H_C$ corresponding to
an energy $E$, then it obeys the differential equation:
\begin{equation}\label{eqHc33}
-\psi''(x)-\frac{2\psi(x)}{|x|}=E\psi(x),\quad x\neq 0.
\end{equation}

In order to get the behavior at the origin of the eigenfunctions of
$H_C$, we integrate by parts and use \eqref{eqHc33}. For
$\varepsilon>0$ we have:
\begin{align}\label{uberkul1}
& \int_{\R}\overline{\phi'(x)}\psi'(x)dx\\
&=\overline{\phi(-\varepsilon)}\psi'(-\varepsilon)-\int_{-\infty}^{-\varepsilon}\overline{\phi(x)}\psi''(x)dx-\overline{\phi(\varepsilon)}\psi'(\varepsilon)-\int_{\varepsilon}^{\infty}\overline{\phi(x)}\psi''(x)dx\nonumber
\\
&+\int_{-\varepsilon}^{\varepsilon}\overline{\phi'(x)}\psi'(x)dx\nonumber
\end{align}
where the last integral will converge to zero as $\varepsilon$ goes to zero. On the other hand,
\begin{align}\label{uberkul2}
& C_0(\phi,\psi)=\int_{-\varepsilon}^0 \ln(-x)\cdot (d_x(\overline{\phi}\psi))(x)dx-\int_0^{\varepsilon} \ln(x)\cdot (d_x(\overline{\phi}\psi))(x)dx\nonumber\\
&+\ln(\varepsilon)\left(\overline{\phi(\varepsilon)}\psi(\varepsilon)+\overline{\phi(-\varepsilon)}\psi(-\varepsilon)\right)+\int_{\R\backslash\left[-\varepsilon,\varepsilon\right]} \frac{1}{|x|}\cdot \overline{\phi(x)}\psi(x)dx.
\end{align}
Then adding \eqref{uberkul1} with \eqref{uberkul2}, using \eqref{eqHc33}
and letting $\varepsilon$ tend to zero (see \cite{CDR} for technical
details) we have:
\begin{align*}
\lim_{\varepsilon\to0}t_C(\phi,\psi)&=\langle \phi,E\psi\rangle +2\lim_{\varepsilon\to0}\overline{\phi(0)}\left[\frac{\psi'(-\varepsilon)-
\psi'(\varepsilon)}{2}+2\ln (\frac{r}{2}) \psi(0)-\ln(\varepsilon)\left(2\psi(0)\right)\right].
\end{align*}
Now using \eqref{centralstation} and the eigenvalue equation
$H_C\psi=E\psi$ we get the boundary condition:
\begin{equation}\label{bdcond}
\lim_{\varepsilon\to0}\left[\frac{\psi'(-\varepsilon)-\psi'(\varepsilon)}{2}+2\ln(\frac{r}{2})\psi(0)-2\ln(\varepsilon)\psi(0)\right]=0.
\end{equation}
\subsection{Eigenvalues and eigenfunctions}
We now have to solve the equation
\begin{equation}\label{theequation}
-\partial_x^2\psi-\frac{2}{|x|}\psi=E\psi,\qquad x\neq0,
\end{equation}
with the boundary condition~\eqref{bdcond}.
Similarly to Loudon in~\cite{L}, we introduce a dimensionless quantity $\alpha$ and the change of variables 
\begin{equation}
E=-\frac{1}{\alpha^2} \mbox{ \rm and } x=\frac{\alpha}{2} z, 
\end{equation}
then we obtain
\begin{equation}\label{eqqquuu}
\frac{d^2}{dz^2}\widetilde{\psi}-\frac{1}{4}\widetilde{\psi}+\frac{\alpha}{|z|}\widetilde{\psi}=0,\qquad z\neq0. 
\end{equation}
The solutions are known for $z>0$ and $z<0$, see~\cite[chap. 13]{AS} for example. The second thing is to see what condition at $z=0$ should the eigenfunctions $\widetilde{\psi}$ obey. If we scale~\eqref{bdcond},
\begin{equation}\label{conditionalfa}
\lim_{\varepsilon\to0}\left[\frac{\widetilde{\psi}'(-\varepsilon)-\widetilde{\psi}'(\varepsilon)}{2}+\alpha(\ln r-\ln(\alpha\varepsilon))\widetilde{\psi}(0)\right]=0.
\end{equation}
We only give here the results. Details of calculations can be found in~\cite{CDR}. If we take the only square integrable solution, we have two cases: If $\alpha=N$ is a positive integer, then the eigenstates are the odd functions $\widetilde{\psi}_{n_{\alpha}p}$ with associated eigenvalues $E_{n_{\alpha}p}$, where $n_{\alpha}=N+1$,
$$\widetilde{\psi}_{n_{\alpha}p}(z)=e^{-\demi |z|}z\frac{1}{\sqrt{2N}}L_{N-1}^1(|z|),\qquad E_{n_{\alpha}p}=-\frac{1}{N^2},
$$
where $L_{N-1}^1$ is an associated Laguerre polynomial. Notice that these energies are independent of the radius of the tube. 
If $\alpha$ is not an integer, the eigenstates are the even functions $\widetilde{\psi}_{n_{\alpha}s}$ with associated eigenvalues $E_{n_{\alpha}s}$, where $n_{\alpha}$ is the smallest integer larger than $\alpha$ and:
$$\widetilde{\psi}_{n_{\alpha}s}=C_{\alpha} W_{\alpha,\frac{1}{2}}(|z|)=C_{\alpha} |z|e^{-\demi |z|}U(1-\alpha,2,|z|)\qquad E_{n_{\alpha}s}=-\frac{1}{\alpha^2},
$$
where $C_{\alpha}$ is a normalizing constant, $W$ is the Whittaker function and $U$ is the confluent hypergeometric function or Kummer function of the second kind. We denote with $\Gamma(z)$ and $\Psi(z)=\Gamma'(z)/\Gamma(z)$ the
usual gamma and digamma functions, and we get from~\eqref{conditionalfa} that for even solutions $\alpha$ must satisfy the relation:
\begin{equation}\label{conditioneven}
\Psi(1-\alpha)+2\gamma+\frac{1}{2\alpha}-\ln\alpha+\ln r=0.
\end{equation}
From this relation, which contains an implicit expression for
$\alpha(r)$, one can deduce several important facts. 
For all integers $N$ and for $\alpha$ in between $N$ and $N+1$, there
is only one solution of~\eqref{eqqquuu} 
satisfying the boundary condition. Furthermore, the energies
associated with the non integer $\alpha$ tend to their closest 
lower integer when $r$ tends to zero. A special case is the one of the
ground state $E_{1s}$ of the exciton which tends to 
minus infinity as the radius tends to zero. The behavior for small $r$ is
$$E_{1s}\stackrel{r\to0}{=}-4(\ln r)^2.
$$

\begin{figure}
\includegraphics[width=3cm]{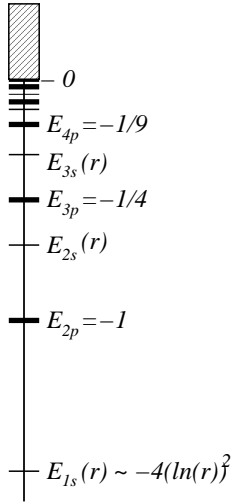}
\caption{Energy of the bound states of the exciton. Energies are expressed in multiples of the effective Rydberg of the exciton.}
\label{figenergy}
\end{figure}

Notice that equation~\eqref{conditioneven} is exactly what one gets by explicitly calculating the condition requiring the derivative of even states to vanish at the origin, relation~(3.22) in Loudon's paper, and replacing his parameter $a$ by $r\cdot a_0/2$, for small $r$. We can also compare with the result from~\cite{PRB87} where Banyai et al. found numerically $a=0.3a_0\cdot r$ for the exciton problem in a quantum wire of radius $r$.
The energies associated to the exciton are drawn on figure~\ref{figenergy} and we calculated numerically the lowest four eigenvalues with respect to the radius on figure~\ref{comparisonfigA}, along with a comparison of the ground state given by the variational method of~\cite{P1}. On figure~\ref{comparisonfigB} a zoom was made around the first and second excited states.

One can notice that the odd states possess a constant energy, independent of the radius. This is due to the fact that the odd states vanish at zero where there is the singularity of the potential. Indeed, the potential becomes, for an odd function $f$:
\begin{equation}
\langle f,V_{\rm eff}^r f\rangle\simeq C_0(f, f),\nonumber
\end{equation}
which is independent of $r$.
\section{Variational approach and comparison}
The variational method operates by minimizing the energy of a trial
function and is therefore usually applied to approximate the ground
state. 
However, by restricting the trial function to forms that are
orthogonal to the ground state, low lying excited states can be
obtained variationally as well. 
In Ref. \cite{P1}, the variational method was applied to the ground
state and in Ref. \cite{P3} 
a similar approach was applied to calculate the $2p$ oscillator
strength of interest for two-photon absorption. 
The $2p$ state is especially important because this state is used for
excitation in two-photon fluorescence measurements \cite{Sc05}. 
By recording the energy of photons emitted from the lowest ($1s$)
exciton, a direct measure of the $2p-1s$ energy difference is
obtained. 
In turn, the $1s$ exciton binding energy (i.e the 1s excitation energy
measured relative to the band gap) can be derived if a 
reliable model for the exciton energy spectrum is invoked.

In the present work, we wish to compare results of the relatively
complicated variational approach to the straight-forward and 
analytical Coulomb model. Hence, in the following we briefly explain
the reasoning behind our variational estimate of the lowest 
excited ($2p$) state. In practical applications, the trial function
must be sufficiently simple that calculation of the energy is
manageable. 
This implies that relatively few adjustable parameters should be
considered. At the same time, a certain flexibility is required to 
provide reasonable accuracy. A useful strategy consists in
constructing trial functions so that they correctly accommodate the
known 
solutions in limiting cases of the general problem. Thus, we are
guided by the analytical solution for the plane, i.e. for nanotube 
radii much larger than the effective Bohr radius. In our units, this
state is simply $\varphi_{2p}\propto x \exp\{-2/3(x^2+y^2)^{1/2}\}$. 
On the other hand, in the extreme 1D limit we expect
$\varphi_{2p}\propto x \exp\{-|x|\}$. To accommodate both limits, we
consequently 
suggest the (un-normalized) trial form
\begin{equation}
\varphi_{2p}(x,y)= x \exp\{-(x^2/k^2+y^2/q^2)^{1/2}\}, 
\end{equation}
where $q$ and $k$ are variational parameters to be determined by
minimizing the expectation value of the energy . This expectation
value is 
found as $E_{2p}=(K-V)/N$, where $K$, $V$ and $N$ are the kinetic
energy, potential energy and normalization constant, respectively. 
The integrations are quite cumbersome and only $K$ and $N$ can be
obtained in closed form in terms of Struve and modified Bessel
functions. 
The potential energy $V$ is evaluated numerically using Gaussian quadrature. 

\begin{figure}
\includegraphics[width=10cm]{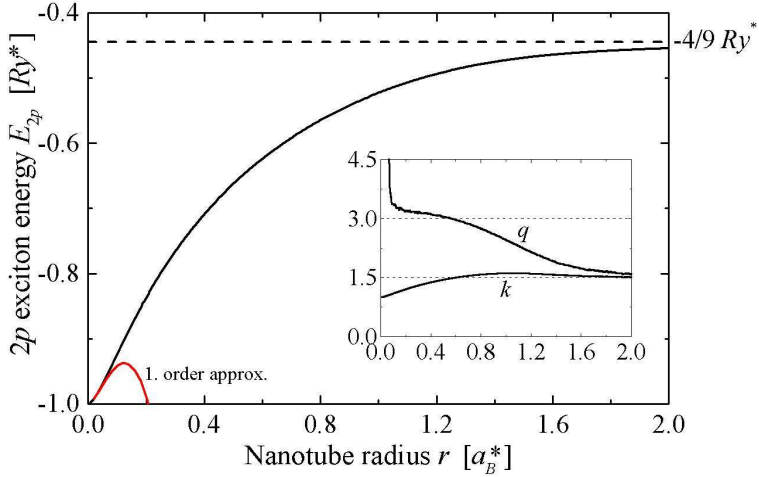}
\caption{Variational $2p$ state energy as a function of nanotube radius. Inset: the $r$-dependence of decay lengths $k$ and $q$ of the trial wave function.}
\label{varia2p}
\end{figure}

The minimized energy as a function of radius $r$ is illustrated in Fig. \ref{varia2p}. 
The limiting values are $E_{2p}=-1$ and $E_{2p}=-4/9$ for small and
large $r$, respectively, in agreement with the analytic solutions in
these limits. 
In between these limits, the curve interpolates smoothly between the limiting values and the dominant correction at small $r$ is $-8(1+\gamma+\ln(r))r^2$.


In order to judge the usefulness of the different approaches used in
the present work it is essential to 
determine the appropriate nanotube radius $r$ in excitonic units,
i.e. in units of $a_B^*$. 
An important point in this respect is that, in fact, $a_B^*$ varies
between different nanotubes. 
The relation needed for conversion is $a_B^*=0.529 \mbox{\rm \AA}
\varepsilon/\mu$, where $\varepsilon$ is 
the dielectric constant screening the Coulomb interaction and $\mu$ is
the reduced effective mass. 
Whereas $\varepsilon$ may be assumed the same for all nanotubes, $\mu$
must be derived from the curvature of the 
band structure and, hence, depends on both radius and chirality of the
nanotube. However, detailed studies \cite{P2} 
show that $a_B^*$ is roughly proportional to the nanotube diameter and
as a consequence $r$ (in units of $a_B^*$) 
is nearly constant and given by $r\approx 0.1 a_B^*$ if
$\varepsilon\approx 3.5$. It should be noted, though, 
that in media with little screening a larger $r/a_B^*$ is
expected. The smallness of $r$ means that approximations based 
on expansion around $r=0$ are expected to be accurate.

\begin{figure}
\includegraphics[width=10cm]{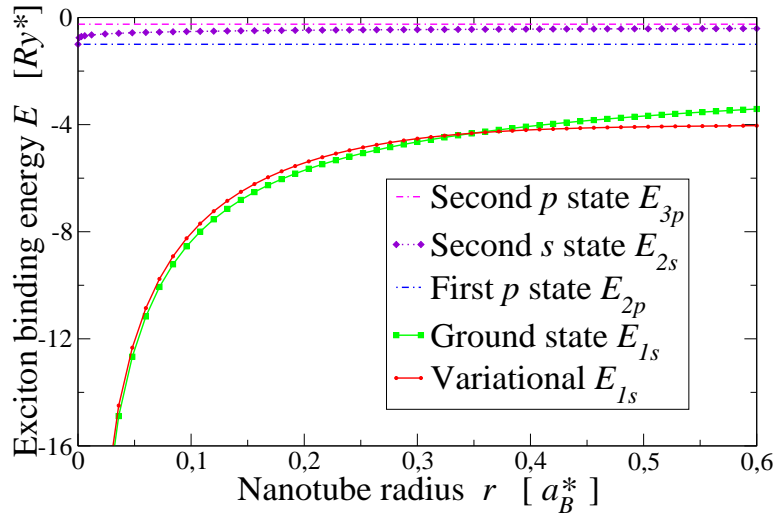}
\caption{Energy of the four lowest bound states of the exciton with
  respect to the radius. Even states were calculated numerically 
using~\eqref{conditioneven}. Notice that the energy of the second $s$
state is equal to the one of the first $p$ state at zero. 
In red: graph of the ground state computed with the variational method on the cylinder from~\cite{P1}.}
\label{comparisonfigA}
\end{figure}

\begin{figure}
\includegraphics[width=10cm]{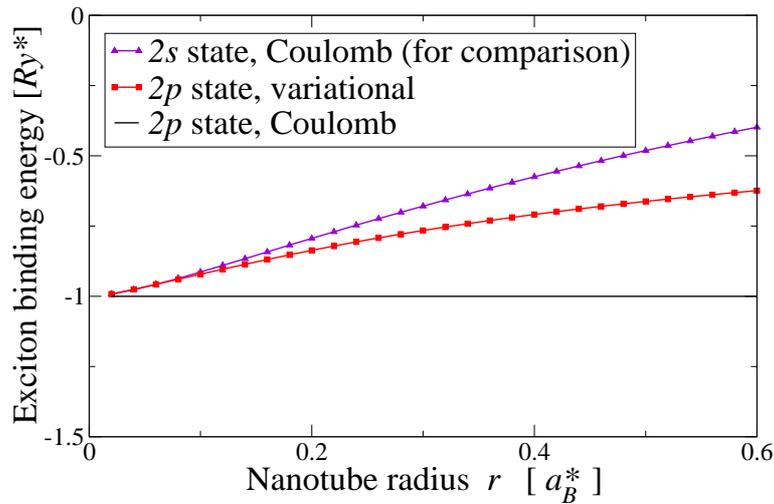}
\caption{Zoom around the energy of the $2p$ state showing both
  variational and Coulomb model results. In addition, 
the $2s$ state obtained with Coulomb model is included for comparison.}
\label{comparisonfigB}
\end{figure}

In figure \ref{comparisonfigA} and \ref{comparisonfigB}, a comparison
of variational energies and the results of the Coulomb model is
given. 
In both the $1s$ and $2p$ cases, reasonable agreement between the two
approaches is found around $r\approx 0.1 a_B^*$. 
Also, in both cases, the variational result lies slightly higher than
the Coulomb model. 
The ground state has deviated by more than $100\%$ from the plane
value for radii around $0.1 a_B^*$. 
An error of less than $5\%$ is seen between the curves at this
point. Note that if the curve from 
the Coulomb model is lower in energy than the variational one, this
does not means that the approximation is better. 
Although in the variational approach the lower is the better since the
exact solution is always below the variational result, 
in the Coulomb model the exact value is somewhere around the
approximation, bounded by an error bound. 
In this work the error bound was not calculated because,
unfortunately, the compromises made to get a 
simple formula implied a difficult calculus to optimize on the bound,
even by numerical computations. For 
the first excited state, the Coulomb model approximation for the
energy is independent of the radius of the tube. 
This must be a good approximation for very small radii. However, this
approximation becomes 
increasingly inaccurate as the radius increases, as the exact value
should tend to the energy of the problem on 
the plane with energy equal to $-4/9a_B^*$ as do the variational curve.  

\section{Conclusion}

The Coulomb model applied to excitons in carbon
nanotubes demonstrates that the energy associated with each even state
decreases to the energy of its closest odd state as the radius tends
to zero. Since each odd state energy is independent of the size of 
the tube in this approach, all energies associated to excited states
stay in the range $(-1,0)$ (in effective Rydberg energy units). 
So only the ground state diverges when the radius gets smaller. This
is confirmed by the behavior of the first variational excited state which
converges to $-1$. 
The simple exciton model proposed in this paper and studied both by
means of rigorous perturbation theory and by a variational approach is
a good starting point for attacking the difficult problem of
electron-electron interactions in low dimensional structures such as
carbon nanotubes.

\section{Acknowledgments.} H.C. acknowledges support from the Danish 
F.N.U. grant {\it  Mathematical Physics and Partial Differential
  Equations}.


\end{document}